\documentclass[a4paper,twocolumn,aps,preprintnumbers,showpacs,superscriptaddress,nofootinbib,amsmath,amssymb]{revtex4}
\usepackage{graphicx}
\usepackage{hyperref}
\usepackage{subfigure}
\usepackage{multirow}
\usepackage[toc,page]{appendix}
\usepackage{color}
\DeclareMathAlphabet{\pazocal}{OMS}{zplm}{m}{n}

\def\imo{i}
\def\re#1{{\rm Re}(#1)}
\def\im#1{{\rm Im}(#1)}

\def\qq{\qquad}

\def\lal{&&\!\! {}}
\def\eq{Eq.\,}
\def\eqs{Eqs.\,}
\def\beq{\begin{equation}}
\def\eeq{\end{equation}}
\def\bear{\begin{eqnarray}}
\def\bearr{\begin{eqnarray} \lal}
\def\ear{\end{eqnarray}}
\def\earn{\nonumber \end{eqnarray}}

\def\yyy{\\[5pt] \lal }

\def\rf{\eqref}
\def\const{{\rm const}}

\begin{document}

\title{Overtones' outburst and Hawking evaporation of Kazakov-Solodukhin quantum corrected black hole}

\author{S. V. Bolokhov}\email{bolokhov-sv@rudn.ru}
\affiliation{Peoples' Friendship University of Russia (RUDN University), 6 Miklukho-Maklaya Street, Moscow, 117198, Russia}
\author{K. A. Bronnikov}\email{kb20@yandex.ru}
\affiliation{Center of Gravitation and Fundamental Metrology, VNIIMS, Ozyornaya ulitsa 46,
	Moscow 119361, Russia}
\affiliation{Peoples' Friendship University of Russia (RUDN University), 6 Miklukho-Maklaya Street, Moscow, 117198, Russia}
\affiliation{National Research Nuclear University ''MEPhI'', Kashirskoe shosse 31,
		Moscow 115409, Russia}
\author{R. A. Konoplya}\email{roman.konoplya@gmail.com}
\affiliation{Institute of Physics and Research Centre of Theoretical Physics and Astrophysics, 		
		Faculty of Philosophy and Science, Silesian University in Opava,
		CZ-746 01 Opava, Czech Republic}

\begin{abstract}
  The Kazakov-Solodukhin black hole metric represents a spherically symmetric deformation
  of the Schwarzschild solution due to quantum-gravity corrections. Assuming the absence of
  nonspherical deformations of the metric, this problem was solved nonperturbatively. In
  this study, we investigate the intensity of Hawking radiation in the background of
  such quantum-corrected black holes and the behavior of their quasinormal modes (QNM). 
  Our findings indicate that while the geometry and such classical characteristics as the 
  fundamental QNM frequencies or the shadow radius are only slightly altered, the Hawking
  radiation and the frequencies of QNM overtones of sufficiently small black holes change 
  much more significantly. This Hawking radiation enhancement arises due to much larger 
  grey-body factors, while the Hawking temperature remains unaffected. 
  The effect becomes significant at the latest stage of black hole evaporation.
\end{abstract}

\maketitle

\section{Introduction}\label{introduction}

  Various approaches to quantization of gravity suggest that the Schwarzschild solution,
  describing a spherically symmetric black hole (BH) in General Relativity, requires
  quantum corrections. Some of these approaches involve finding quantum-deformed BH
  geometries through semiclassical corrections caused by vacuum polarization
  due to matter fields near a BH \cite{York,Page} or even deeply inside its horizon  
  \cite{we1, we2}.The proper oscillation frequencies of semiclassically corrected BHs
  were analyzed in \cite{Piedra:2009pf,Piedra:2010ur} based on this approach. However, 
  this method does not take into account the dominant factor, which is quantization of the
  gravitational field itself. Fortunately, if we neglect nonspherical deformations
  and use effective scalar-tensor gravity, the problem becomes renormalizable. Kazakov
  and Solodukhin found the corresponding generalization of the Schwarzschild solution long
  ago \cite{Kazakov:1993ha}. The thermodynamic properties of such BHs were analyzed
  in \cite{Kim:2012cma,Shahjalal:2019ypz,Lobo:2019put}, and some cosmological applications
  were discussed in \cite{Shahjalal:2019fig}. Quasinormal modes (QNM) of these BHs were
  considered in \cite{Saleh:2016pke,Saleh:2014uca,Konoplya:2019xmn}, and the
  grey-body factors for the electromagnetic field in \cite{Konoplya:2019xmn}. Some
  generalizations of the Kazakov-Solodukhin space-time and their thermodynamic properties
  were considered in \cite{Simpson:2021dyo}. Optical properties during the accretion process
  in the vicinity of Kazakov-Solodukhin BHs were studied in \cite{Huang:2023ilm}.

  Here, we are interested in two thus far unexplored consequences of these quantum 
  corrections on the Hawking radiation of the Kazakov-Solodukhin BHs and the frequencies of
  their QNM overtones. One should note here that the Hawking formula was obtained in 
  a semiclassical setup, in which the background metric is still classical but the matter fields 
  are quantum. Our consideration here has a hybrid nature: we still treat the Hawking radiation
  via Hawking's well-known formula, but take into consideration the dominant effective
  corrections to the BH metric due to quantum properties of the gravitational field. As the
  resulting metric is still static and spherically symmetric, one can expect that this approach
  will be reasonable for most of the evaporation time, while the BH mass is well above the 
  Planck mass. A firm reason for this expectation is that, as we know from Page's seminal work
  \cite{Page:1976df}, the Hawking radiation in the form of gravitons contributes only about 
  $2\%$ of the total emission of all particles. Therefore, once we are interested in the 
  emission of matter fields, we can use the Hawking formula since these matter fields do not 
  care of the origin of the static metric and its possible quantum-gravity corrections. One more
  argument in favor of this semiclassical approximation up to the latest stage of BH evaporation 
  is that the strong quantum--gravity corrections of the metric are only relevant in a very small
  region of Planck size, many orders of magnitude smaller than the Compton wavelength of
  any Standard model particle. 
  On the other hand, since the effect under study must come into force not too far from 
  the Planck energy scale, hence at energies many orders of magnitude larger than the 
  particle masses, all Standard Model particles can be regarded ultrarelativistic, and it is 
  reasonable to treat them as massless ones in the calculations.

  To study the evaporation of Kazakov-Solodukhin BHs, we re-examine their grey-body factors 
  by adding the relevant calculations for the scalar, Maxwell and Dirac fields with various
  multipole numbers, which is necessary for obtaining the intensity of Hawking radiation of
  all Standard Model fields, while earlier only Maxwell field data were considered for the
  first two multipoles \cite{Konoplya:2019xmn}. We show that while the quantum correction
  changes the BH geometry relatively weakly, leading, for example, to a few percent change of 
  the fundamental QNM, the Hawking radiation intensity of sufficiently small BHs can change 
  as much as by tens of percent.
   
  The organization of the paper is as follows: Section II provides basic information about
  Kazakov-Solodukhin BHs. In Section III, we discuss the scattering formalism and
  grey-body factors for the scalar, Maxwell and Dirac fields with various multipole numbers.
  In Section IV, we estimate the intensity of Hawking radiation of Kazakov-Solodukhin
  BHs and compare them with the Schwarzschild case. Section V is devoted to QNM 
  overtones of Kazakov-Solodukhin BHs and a comparison of their frequencies with those
  of Schwarzschild  BHs. Finally, in the Conclusion
  we summarize the results and mention some open questions.

\section{The Kazakov-Solodukhin metric and wave equations}

  The deformation of the Schwarzschild solution of general relativity due to spherically symmetric
  quantum fluctuations of the metric was obtained by Kazakov and Solodukhin in \cite{Kazakov:1993ha}.
  In that case, the 4D theory of gravity with Einstein action reduces to effective 2D dilaton gravity.
  The Kazakov-Solodukhin metric \cite{Kazakov:1993ha} has the form
\begin{equation}
		d s^2 = - f(r) d t^2 + f^{-1}(r) d r^2 + r^2 (d \theta^2 + \sin^2 \theta d \varphi^2),
\end{equation}
  where the quantum-gravity corrected function $f(r)$ is
\bearr       \label{2}
       f(r) =  \frac{\sqrt{r^2-a^2}}{r}-\frac{2 M}{r}, \qq  a = \const > 0,
\ear
  $a$ being a deformation parameter. At $a = 0$, this metric reduces to the Schwarzschild one,
  but unlike the latter, it is not Ricci-flat, and the Ricci scalar is
\begin{equation}\label{eq:R}
		R(r) = \frac{2}{r^2} \left[1 - \bigg(1-\frac{a^2}{r^2}\bigg)^{\!\! -1/2}\right]
				+\frac{a^2}{r^4} \left(1-\frac{a^2}{r^2} \right)^{\!\! -3/2}.
\end{equation}
  The event horizon $r_h$ is situated at
$$
		r_h =\sqrt{4 M^2 + a^2}.
$$
  The singularity is located at $r =a$ since as $r\to a$ the curvature tends to infinity.

  The neutral massless scalar field $\Phi$ obeys the covariant d'Alembert equation
\begin{equation}\label{K-G}
    \frac{1}{\sqrt{-g}}\partial _{\mu }\left(g^{\mu\nu}\sqrt{-g}\partial_{\nu}\Phi \right)=0.
\end{equation}

  The electromagnetic field equation has the form
\begin{equation}\label{EmagEq}
		\frac{1}{\sqrt{-g}}\partial_\mu \left(F_{\rho\sigma}g^{\rho \nu}g^{\sigma \mu}\sqrt{-g}\right)=0\,,
\end{equation}
  where $F_{\rho\sigma}=\partial_\rho A_{\sigma}-\partial_\sigma A_{\rho}$, and $A_\mu$ is the
  vector potential.

  The massless Dirac equation in curved space-time is
\begin{equation}\label{dirac}
            [\gamma^ae^\mu_a(\partial_\mu+\Gamma_\mu)]\Psi=0,
\end{equation}
  where $\gamma^a$ are the Dirac matrices, $e^\mu_a$ is the inverse of the tetrad $e^a_\mu$
  ($g_{\mu\nu} = \eta_{ab}e^a_\mu e^b_\nu$), $\eta_{ab}$ is the Minkowski metric.
  The spin connections $\Gamma_\mu$ are defined as follows:
\begin{equation}
		\Gamma_\mu=\frac{1}{8}[\gamma^a, \gamma^b]e^\nu_ae_{b\nu;\mu}.
\end{equation}

\begin{figure}
\centerline{\resizebox{\linewidth}{!}{\includegraphics*{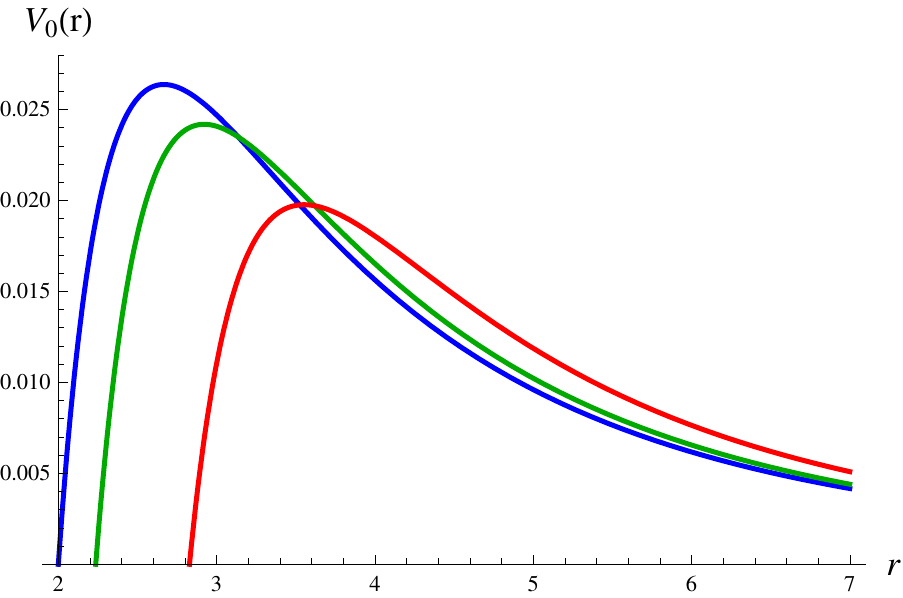}}}
\caption{The effective potential $V_0(r)$ for $\ell=0$ scalar perturbations of 
		Kazakov-Solodukhin BHs with $a=0$ (the Schwarzschild case), $a=1$, $a=2$ (left to right).}
\label{figVscal}
\end{figure}

\begin{figure}
\centerline{\resizebox{\linewidth}{!}{\includegraphics*{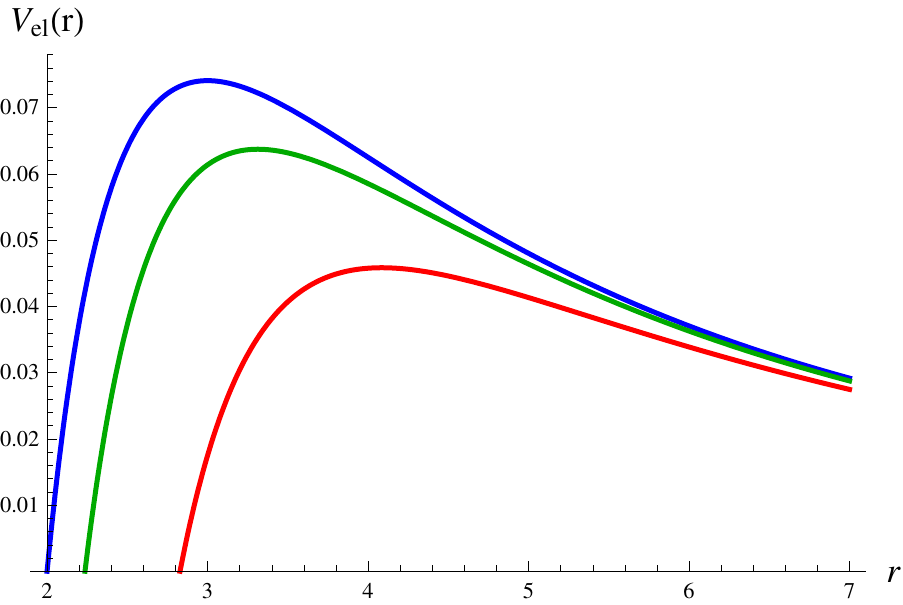}}}
\caption{The effective potential $V_{\rm el}(r)$ for $\ell=1$ electromagnetic perturbations
		of Kazakov-Solodukhin BHs with $a=0$ (the Schwarzschild case), $a=1$, $a=2$ (left to right).}
	 \label{figVem}
\end{figure}

\begin{figure}
\centerline{\resizebox{\linewidth}{!}{\includegraphics*{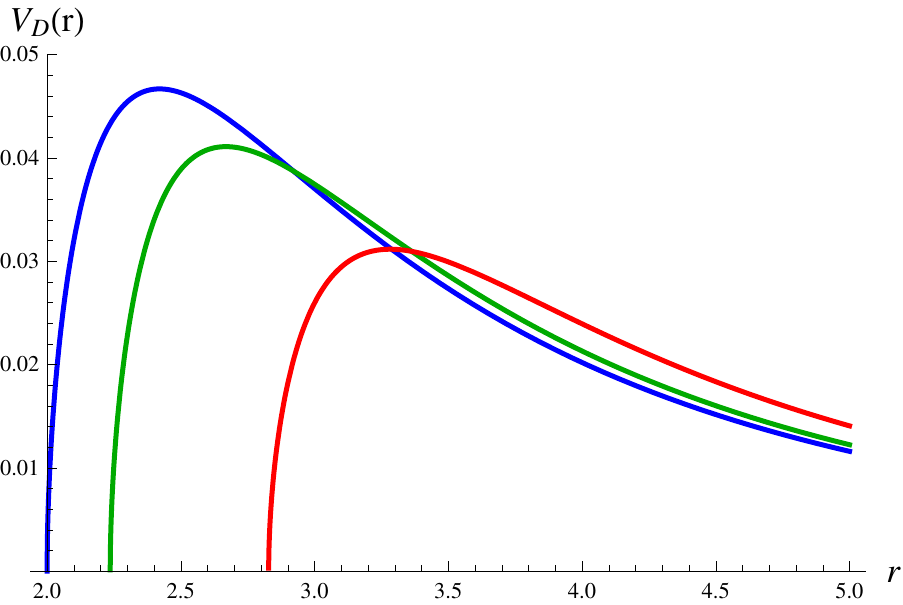}}}
\caption{The effective "plus-type" potential $V_{\rm D}(r)$ for $\ell=1$ 
	Dirac perturbations of Kazakov-Solodukhin BHs with $a=0$ (the Schwarzschild case), $a=1$, $a=2$ (left to right).}
	\label{figVDir}
\end{figure}

   After separation of variables, Eqs.~\eqref{K-G}--\eqref{dirac} take the general
   wavelike form
\begin{equation}\label{wave-equation}
		\frac{d^2 \Psi_s}{dr_*^2} + \left(\omega^2-V_{s}(r)\right) \Psi_s = 0,
\end{equation}
  where  $s=0$ corresponds to the scalar field, $s=\pm 1/2$ to the Dirac field, and $s=1$ to
  the electromagnetic field. The dynamics of the two-component spinor field is reduced to a 
  pair of master wave equations with effective potentials $V_{\pm 1/2}$ \cite{Brill:1957fx}.  
  The ``tortoise coordinate'' $r_*$ is defined as $ dr_* = dr/f(r)$,
  and the effective potentials are
\bearr	  \label{scalarpot}
V_0(r)=f(r) \left(\frac{\ell(\ell+1)}{r^2} + \frac{ f'(r)}{r}\right),
\yyy
                 \label{empotential}
		V_{1}(r)  =  f(r) \frac{\ell (\ell+1)}{r^2},
\yyy
       V_{\pm 1/2} = \frac{\sqrt{f}|\ell|}{r^2}\Big(|\ell|\sqrt{f} 
                     \pm \frac{r}{2}\frac{df}{dr} \mp f\Big),
\ear
  where $|\ell| = 1, 2, 3, \ldots$ is the total angular momentum number.\footnote
  		{{In the Dirac field case, in order to unify the notation, by the symbol $\ell$ 
  		we mean an integer connected with the half-integer fermionic multipole number 
  		$l$ as $\ell = l+1/2$.}} 
  For generic spherically symmetric space-times, the two effective potentials $V_{+1/2}$ 
  and $V_{-1/2}$ are isospectral. Therefore, we will consider only one of them, with the plus sign.
  Examples of the effective potentials are shown in Figs.\,\ref{figVscal}--\ref{figVDir}.

  It is of utmost importance that in the perturbative calculation of the quantized gravitational 
  field leading to the deformed metric, the values of $a$ need not be small 
  \cite{Kazakov:1993ha}, and so the quantum correction can be large. Following 
  the Kazakov-Solodukhin arguments  (see p.\,162 in \cite{Kazakov:1993ha}), we suppose 
  that the minimum radius $a$ can be of the order of the Planck length (formally equal to 
  half the Schwarzschild radius of a BH of Planck mass $M_{\rm Pl}$) but can substantially 
  exceed this minimum value. Thus, fixing the BH mass as $M=1$, we will assume small 
  or moderate deformations of Schwarzschild space-time, such that $a/M\lesssim 5$ for 
  masses not too much greater than the Planck one, say, $M\lesssim 100\, M_{\rm Pl}$.

\section{Scattering problem}

  Grey-body factors determine the proportion of the initial quantum radiation that is reflected 
  back to the event horizon by the potential barrier in its vicinity. Thus we apply
  Hawking's semiclassical formula with a grey-body factor to estimate the amount of radiation
  that will reach a distant observer, considering late phases of BH evaporation and
  using the deformed metric \rf{2} despite its quantum-gravity origin.

  As mentioned above, there are reasons to neglect the radiation of gravitons: as is known using
  Schwarzschild BHs as an example, in the semiclassical domain, gravitons account for less
  than $2\%$ of the total radiation flux, as illustrated in \cite{Page:1976df} and summarized in
  Table I of \cite{Konoplya:2019ppy}. As a result, the grey-body factors of test fields characterize
  not only classical scattering but also the Hawking radiation intensity. Furthermore, as stated in
  \cite{Konoplya:2019ppy}, grey-body factors can even be more influential than the temperature.

  We will examine the wave equation (\ref{wave-equation}) under boundary conditions that allow
  for incoming waves from infinity. Due to the symmetry of scattering properties, this is 
  equivalent to scattering of a wave originating from the horizon. The boundary conditions 
  for scattering in (\ref{wave-equation}) are as follows:
\begin{equation}\label{BC}
\begin{array}{ccll}
    \Psi &=& e^{-i\omega r_*} + R e^{i\omega r_*},& r_* \to +\infty, \\[5pt]
    \Psi &=& T e^{-i\omega r_*},& r_* \to -\infty, \\
\end{array}
\end{equation}
   where $R$ and $T$ are the reflection and transmission coefficients, respectively.

  The effective potential has the form of a potential barrier that decreases monotonically 
  towards both infinities, allowing for application of the WKB approach
  \cite{Schutz:1985zz,Iyer:1986np,Konoplya:2003ii}
  to determine $R$ and $T$. As $\omega^2$ is real, the first-order WKB values for $R$ and 
  $T$ will be real as well \cite{Schutz:1985zz,Iyer:1986np,Konoplya:2003ii}, and
\begin{equation}\label{1}
		|T|^2 + |R|^2 = 1.
\end{equation}
  Once the reflection coefficient is obtained, we can determine the transmission coefficient for each
  multipole number $\ell$,
\begin{equation}
		|{\pazocal A}_{\ell}|^2 = 1 - |R_{\ell}|^2 = |T_{\ell}|^2,
\end{equation}
 where ${\pazocal A}_{\ell}$ is the grey-body factor.

  Several methods are available in the literature for computing the reflection and transmission
  coefficients. In our study, we utilized the higher order WKB formula \cite{Konoplya:2003ii} for 
  relatively accurate estimation of these coefficients. However, this formula is not suitable for 
  very small values of $\omega$, which correspond to almost complete wave reflection and 
  have negligible contributions to the overall energy emission rate. For this mode, we employ
  an extrapolation of the WKB results at a given order to smaller $\omega$. According to 
  \cite{Schutz:1985zz,Iyer:1986np}, the reflection coefficient can be expressed as 
\begin{equation}\label{moderate-omega-wkb}
			R = (1 + e^{- 2 i \pi K})^{-1/2},
\end{equation}
  where $K$ is determined by solving the equation
\begin{equation}
		K - i \frac{(\omega^2 - V_{\max})}{\sqrt{-2 V''_{\max}}}
				- \sum_{i=2}^{i=6} \Lambda_{i}(K) =0,
\end{equation}
  involving the maximum of the effective potential $V_{\max}$, its second-order derivative
  $V''_{\max}$ with respect to the tortoise coordinate, and higher order WKB corrections 
  $\Lambda_i$. Strictly speaking, the WKB series does not guarantee convergence in each order, 
  but only asymptotically, so that usually there is an optimal moderate order at which the accuracy
  is the best. This order depends on the form of the effective potential. Here we used the 6th order
  for Maxwell perturbations and the 3rd order for the Dirac field with a plus-potential, because
  these orders provide the best accuracy in the Schwarzschild limit, and we expect that this will 
  also take place for a deformed BH. This approach was used in a few papers 
  \cite{Konoplya:2019hlu,Konoplya:2010vz,Konoplya:2020cbv,Bolokhov:2024voa}, showing good
  concordance with numerical integration.

\begin{figure}
\centerline{\resizebox{\linewidth}{!}{\includegraphics*{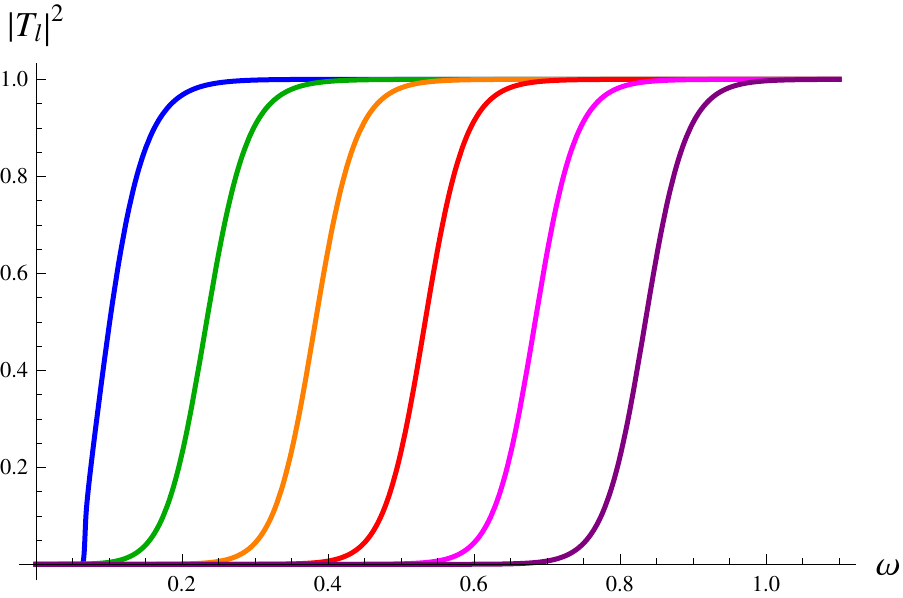}}}
\caption{Grey-body factors of the scalar field per frequency unit for the Kazakov-Solodukhin
	BH with $a=2$ for $\ell=0,1,2,3,4,5$ (left to right).}\label{ScalGBa2}
\end{figure}

\begin{figure}
\centerline{\resizebox{\linewidth}{!}{\includegraphics*{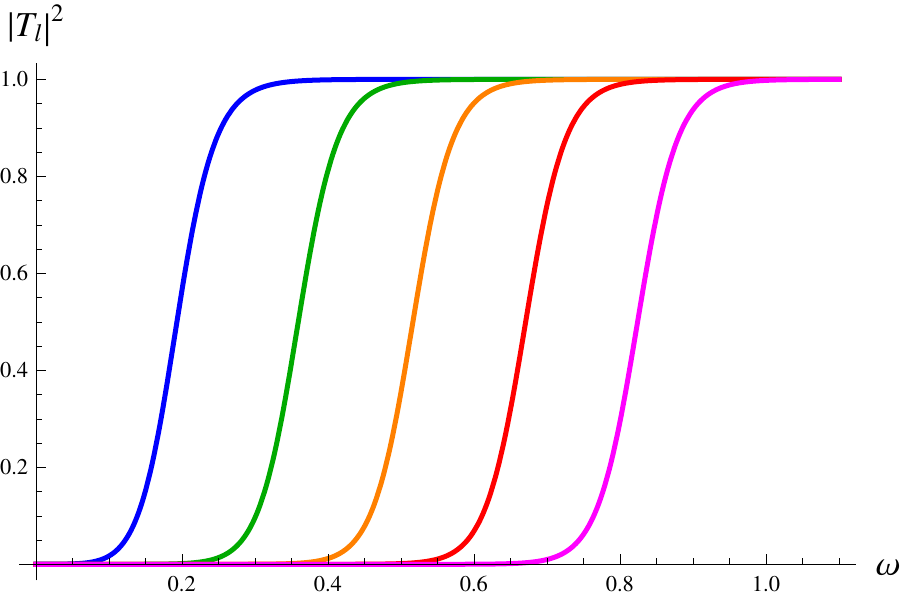}}}
\caption{Grey-body factors of the Maxwell field per frequency unit for the Kazakov-Solodukhin
	BH with $a=2$ for $\ell=1,2,3,4,5$ (left to right). }\label{ElmagGBa2}
\end{figure}

\begin{figure}
\centerline{\resizebox{\linewidth}{!}{\includegraphics*{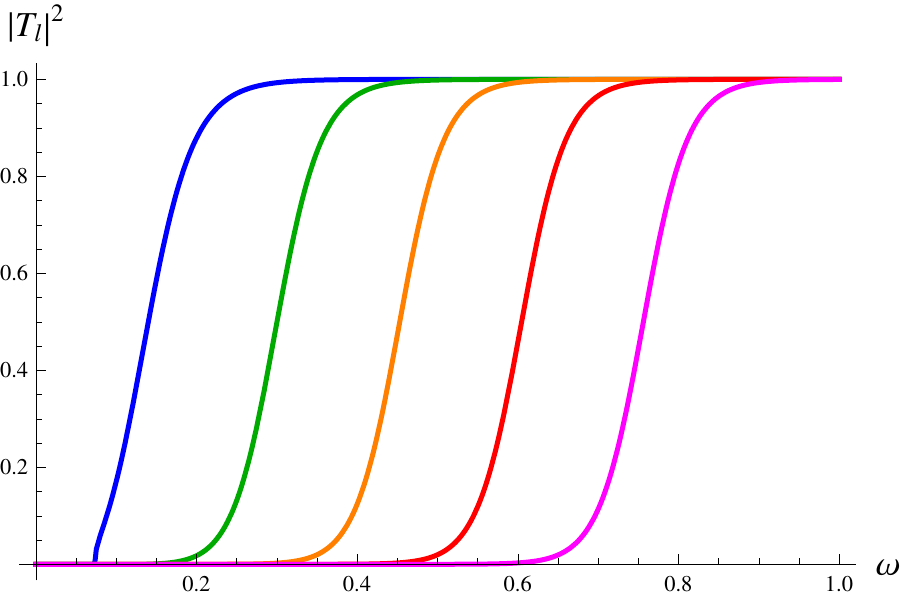}}}
\caption{Grey-body factors of the Dirac field per frequency unit for the Kazakov-Solodukhin
	BH with $a=2$ for $\ell=1,2,3,4,5$ (left to right). }\label{DirGBa2}
\end{figure}

  In the previous work \cite{Konoplya:2019xmn}, the grey-body factors for the Maxwell field were
  calculated. Now we complement that work by determining these factors for the
  scalar and Dirac fields, see Figs.~\ref{ScalGBa2}--\ref{DirGBa2} as examples for the 
  case $a=2$ and various $\ell$. With grey-body factors of these fields at hand, we 
  are ready to find the energy emission rate of the Hawking radiation.


\begin{figure}
\centerline{\resizebox{\linewidth}{!}{\includegraphics*{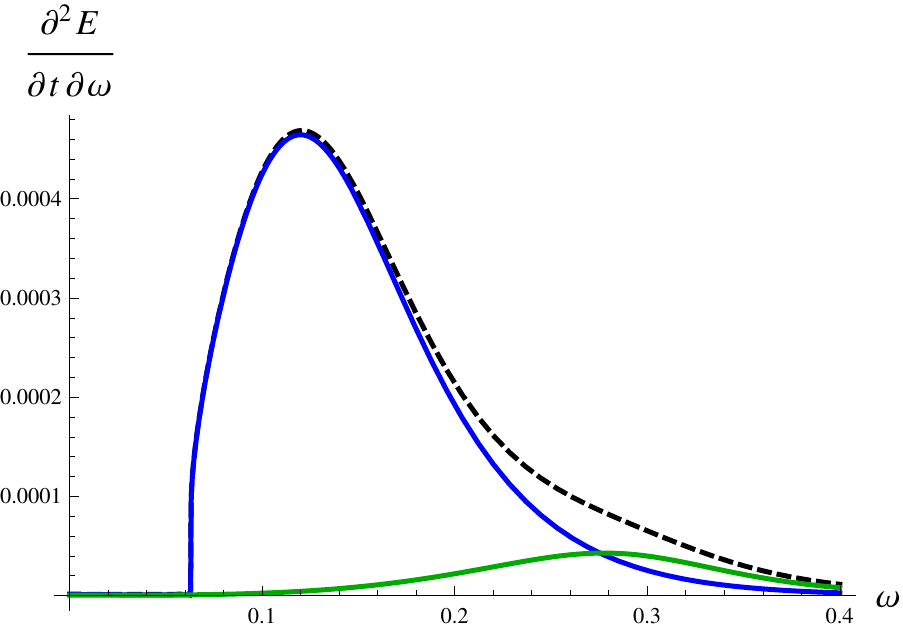}}}
\caption{Energy emission rate of the scalar field per frequency unit for the Schwarzschild case
	$a=0$ and for various values of $\ell$: black dashed curve (top, total), blue curve (middle, $\ell = 0$), green curve (bottom, $\ell =1$). 	
	Emission rates for higher	$\ell$ are strongly suppressed and would be invisible in the plot.}
\label{ScalRate0}
\end{figure}

\begin{figure}
\centerline{\resizebox{\linewidth}{!}{\includegraphics*{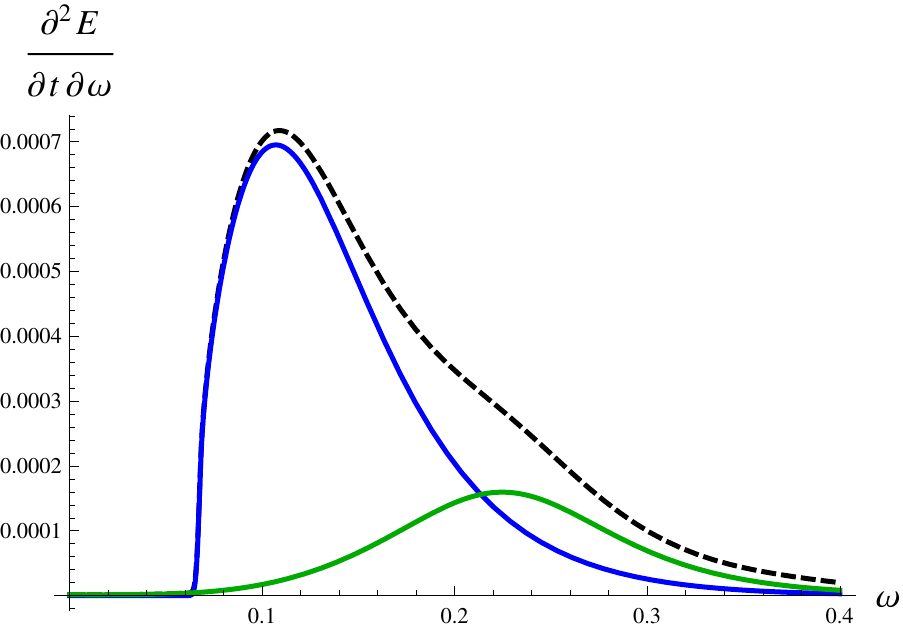}}}
\caption{Energy emission rate of the scalar field per frequency unit for Kazakov-Solodukhin 
	BHs with $a=2$ and for various values of $\ell$: black dashed curve (top, total), blue curve (middle, $\ell = 0$), green curve (bottom, $\ell =1$). Emission rates for higher $\ell$ are strongly suppressed and would be 
	invisible in the plot.}
\label{ScalRate2}
\end{figure}
\begin{figure}
\resizebox{\linewidth}{!}{\includegraphics*{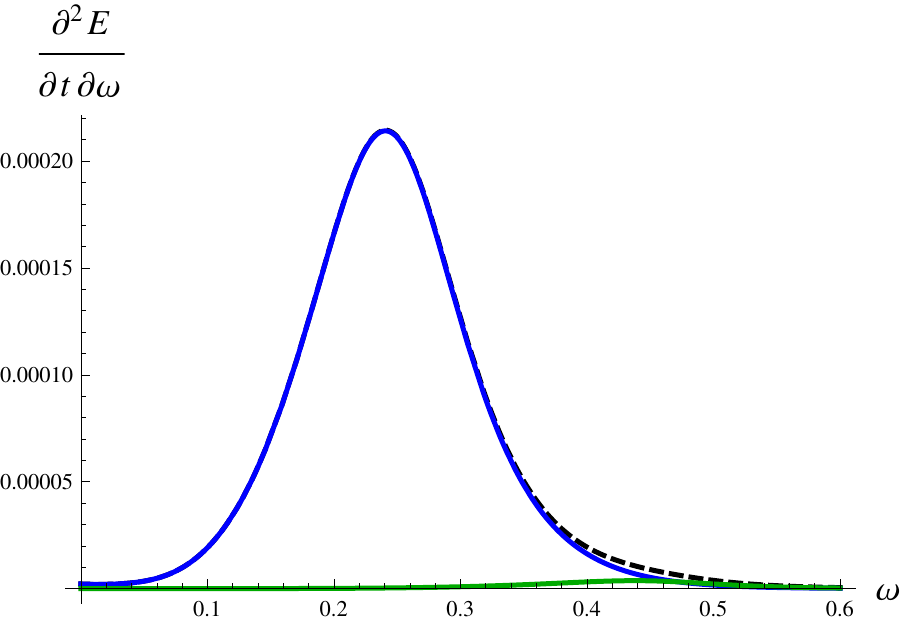}}
\caption{Energy emission rate of the Maxwell field per frequency unit for the Schwarzschild
	case $a=0$ for various values of $\ell$: black dashed curve (top, total), blue curve (middle, $\ell = 1$), green curve (bottom, $\ell =2$).
	 Emission rates for higher $\ell$ are strongly suppressed and would be invisible in the plot.
	}\label{fig41}
\end{figure}
\begin{figure}
\resizebox{\linewidth}{!}{\includegraphics*{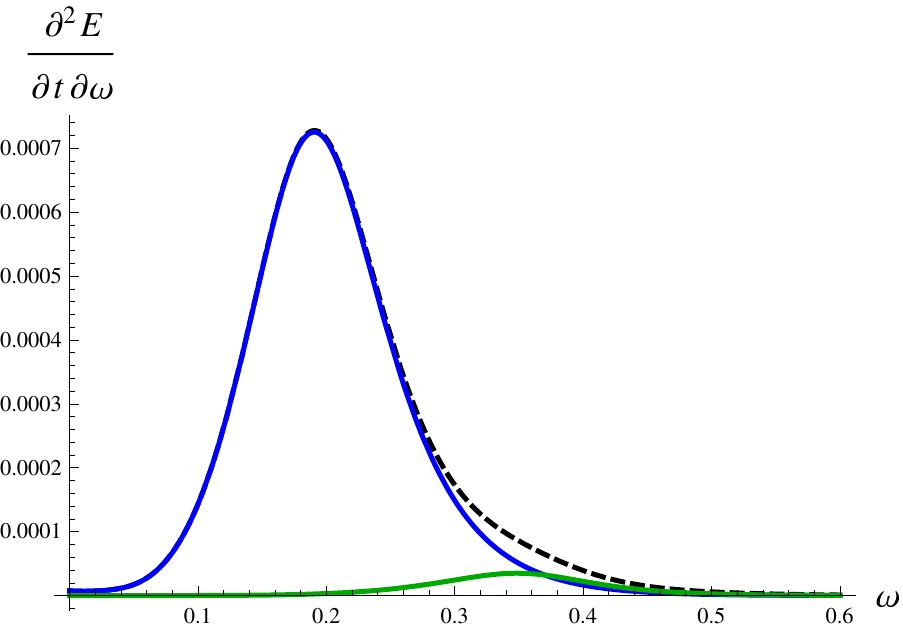}}
\caption{Energy emission rate of the Maxwell field per frequency unit for Kazakov-Solodukhin
	BHs with $a=2$ for various values of $\ell$: black dashed curve (top, total), blue curve (middle, $\ell = 1$), green curve (bottom, $\ell =2$). Emission rates for higher $\ell$ are strongly suppressed
     and would be invisible in the plot.}
\label{fig42}
\end{figure}
\begin{figure}
\centerline{\resizebox{\linewidth}{!}{\includegraphics*{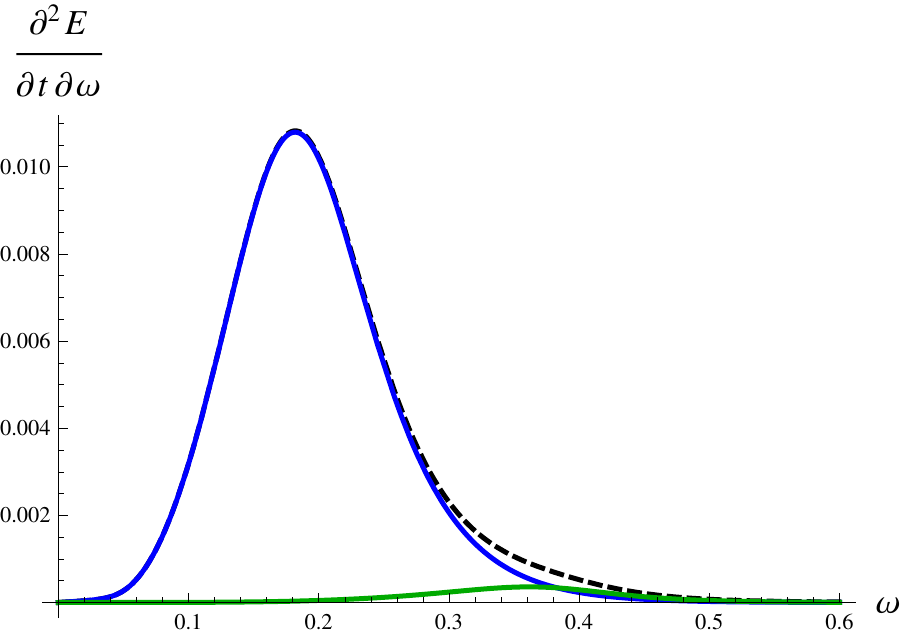}}}
\caption{Energy emission rate of the Dirac field per frequency unit for the Schwarzschild case
	$a=0$ and for various values of
	$\ell$: black dashed curve (top, total), blue curve (middle, $\ell = 1$), green curve (bottom, $\ell =2$). Emission rates for higher
	$\ell$ are strongly suppressed and would be invisible in the plot.}
\label{fig51}
\end{figure}
\begin{figure}
\centerline{\resizebox{\linewidth}{!}{\includegraphics*{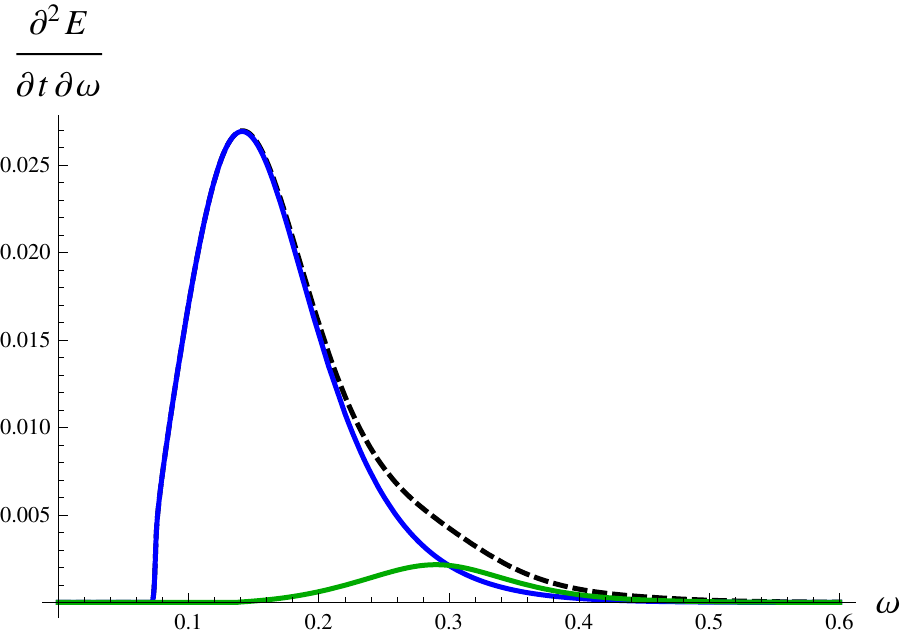}}}
\caption{Energy emission rate of the Dirac field per frequency unit for Kazakov-Solodukhin 
	BHs with $a=2$ and various values of
	$\ell$: black dashed curve (top, total), blue curve (middle, $\ell = 1$), green curve (bottom, $\ell =2$). Emission rates for higher
	$\ell$ are strongly suppressed and would be invisible in the plot.
}\label{fig52}
\end{figure}

\section{Hawking radiation}

  In the following analysis, we make the assumption that the BH is in a state of thermal
  equilibrium with its surroundings. This means that the BH temperature remains constant
  between the emission of two consecutive particles. According to this assumption, the system 
  can be characterized by the canonical ensemble, extensively discussed in the literature 
  (see, e.g., \cite{Kanti:2004nr}). Consequently, the well-known formula for the energy 
  emission rate of Hawking radiation \cite{Hawking:1974sw} can be applied:
\begin{align}\label{energy-emission-rate}
		\frac{\text{d}E}{\text{d} t} = \sum_{\ell}^{} N_{\ell}
		| \pazocal{A}_\ell |^2 \frac{\omega}		
				{\exp\left(\omega/T_\text{H}\right)\pm1}\cdot \frac{\text{d} \omega}{2 \pi},
\end{align}
  where $T_{\rm H}$ is the Hawking temperature, $\pazocal{A}_\ell$ are grey-body factors, 
  and $N_\ell$ are the multiplicities which depend on the number of species of particles and $\ell$.
  The Hawking temperature is \cite{Hawking:1974sw}
$$
		T = \left.\frac{f'(r)}{(4 \pi)}\right\vert_{r=r_{H}}  = \frac{1}{8 \pi  M},
$$
  which is in our case remarkably the same as for a Schwarzschild BH, independent from the
  deformation parameter $a$. The latter means that a deviation of Hawking evaporation 
  from the Schwarzschild case will be entirely determined by the BH grey-body factors.
  

  We are considering BHs large enough for using the semiclassical approximation but small
  enough for the Kazakov-Solodukhin quantum correction to be significant, say, with
  $5 \lesssim M/M_{\rm Pl} \lesssim 100$, hence we can suppose that any massive 
  Standard Model particles are emitted in an ultrarelativistic regime, roughly at the 
  same rate as massless ones, for example, any Dirac particles are emitted roughly at the 
  same rate as massless neutrinos.
  Also, as was argued above, we neglect the emission of gravitons.

  The multiplicity factors for the 4D spherically symmetrical black holes consist of the number 
  of degenerate $m$-modes ($m = -\ell, -\ell+1, ....-1, 0, 1, ...\ell$, that is  $2 \ell +1$ modes)
  multiplied by the number of species of particles, which in turn also depends on the number 
  of polarizations and helicities of particles. Therefore, we have 
\bearr
       	N_{\ell} = 2 \ell+1 \qquad \quad \!(\rm scalar),
\\ \lal
		N_{\ell} = 2 (2 \ell+1) \qquad \!\!\!(\rm Maxwell),
\\ \lal
		N_{\ell} = 84 \ell \qquad \qquad (\rm Dirac), \label{multiDir}
\ear
  The multiplicity factor for the Dirac field is fixed taking into account the degrees of 
  freedom for fermions of the Standard Model (quarks and leptons) in the ultrelativistic regime, 
  including their antiparticles, flavor or generation, polarization or helicity, both the ``plus'' 
  and ``minus'' potentials related by Darboux transformations, leading to isospectral problems, 
  and the same grey-body factors for both chiralities. We are using the ``plus'' potential 
  because the WKB results are more accurate for that case in the Schwarzschild limit.

  The energy emitted makes the BH mass decrease at the following rate \cite{Page:1976df}:
\begin{equation}        \label{M-dot}
		\frac{d M}{d t} = -\frac{\hbar c^4}{G^2} \frac{\alpha_{0}}{M^2}
		\approx - 1.91\times 10^{29}\alpha_0\ \frac{\rm g}{\rm s}
			\cdot\bigg(\frac{\rm g^2}{M^2}\bigg),
\end{equation}
  where we have restored the dimensional constants. Here $\alpha_{0} = d E/d t$ is taken
  in units used in Table I for a given initial mass $M_0$.  Indeed, this can be seen by comparison of
  our data in the Schwarzschild limit with \cite{Page:1976df}, restricting the data to the
  sector of particles which was considered by Page. In this case, a relatively small difference with
  Page's results ($\lessapprox 1.6 \%$) is due to the systematic error of the WKB method
  \cite{Konoplya:2019hlu}.  From Figs.~\ref{ScalRate0}--\ref{fig52} one can see that as $\ell$ 
  grows, the contribution to the Hawking radiation rapidly weakens, so that effectively a few
  first multipoles are already sufficient for a very good estimation of the evaporation. This 
  happens because the potential barrier grows as $\sim \ell^2 +\ell$ and reflects almost the
  whole flux emitted at high multipole numbers.
  
\begin{table}
\centering
\caption{
	The total energy emission rate $dE/dt$ for scalar, Maxwell and Dirac particles taken 
	with appropriate multiplicity factors for various values of the deformation parameter $a$. 
	Summing over $\ell$ is done for the first five multipole moment values. The power is in units 
     $\hbar c^{6} G^{-2} M^{-2}= 1.719 \cdot 10^{50} (M/\rm g)^{-2}\,\rm erg \cdot s^{-1}.$}
\medskip     
\begin{tabular}{|c|c|c|c|}
  \hline
  $a/M$ & Scalar & Maxwell & Dirac   \\
\hline
  0 ($\ell=0$) & $0.0000576251$ &  $-$ & $-$  \\
  0 ($\ell=1$) & $7.27206\times 10^{-6}$ &  $0.0000335107$ &  $0.00175439$  \\
  0 ($\ell=2$) & $1.93201\times 10^{-7}$ &  $6.67916 \cdot 10^{-7}$ &  $0.0000616244$  \\
  0 ($\ell=3$) & $3.34703\times 10^{-9}$ &  $1.00693 \cdot 10^{-8}$ &  $1.22529\times 10^{-6}$ \\
\hline
  0 (total) & $0.0000650938$ & $0.0000341888$  &  $0.00181726$ \\
\hline
  0.5 ($\ell=0$) & $0.0000591509$ &  $-$ & $-$  \\
  0.5 ($\ell=1$) & $8.14123\times 10^{-6}$ &  $0.0000371931$ &  $0.00179025$ \\
  0.5 ($\ell=2$) & $2.37239\times 10^{-7}$ &  $8.16104 \cdot 10^{-7}$ &  $0.0000720743$ \\
  0.5 ($\ell=3$) & $4.51999\times 10^{-9}$ &  $1.35512 \cdot 10^{-8}$ &  $1.57593\times 10^{-6}$\\
\hline
  0.5 (total) & $0.0000675339$ &   $0.000038023$  &  $0.00186393$  \\
\hline
  1.0 ($\ell=0$) & $0.0000628496$ &  $-$ & $-$  \\
  1.0 ($\ell=1$) & $0.0000109211$ &  $0.0000487351$ &  $0.00212684$ \\
  1.0 ($\ell=2$) & $4.04667\times 10^{-7}$ &  $1.37279 \cdot 10^{-6}$ &  $0.000108223$  \\
  1.0 ($\ell=3$) & $9.87953\times 10^{-9}$ &  $2.93303 \cdot 10^{-8}$ &  $3.03148\times 10^{-6}$  \\
\hline
  1.0 (total) & $0.0000741855$ &  $0.0000501377$ &  $0.00223816$  \\
  \hline
  1.5 ($\ell=0$) & $0.0000676597$ &  $-$ & $-$  \\
  1.5 ($\ell=1$) & $0.0000160219$ &  $0.0000692847$ &  $0.00265984$  \\
  1.5 ($\ell=2$) & $8.10836\times 10^{-7}$ & $2.69491 \cdot 10^{-6}$ &  $0.000183381$  \\
  1.5 ($\ell=3$) & $2.73734\times 10^{-8}$ &  $8.00841 \cdot 10^{-8}$ &  $7.09824\times 10^{-6}$  \\
\hline
  1.5 (total) & $0.0000845207$ & $0.0000720618$ &  $0.00285053$  \\
  \hline
  2.0 ($\ell=0$) & $0.0000747864$ &  $-$ & $-$  \\
  2.0 ($\ell=1$) & $0.0000238968$ & $0.000100234$ & $0.00335062$  \\
  2.0 ($\ell=2$) & $1.66429\times 10^{-6}$ & $5.39958 \cdot 10^{-6} $ & $0.000315782$  \\
  2.0 ($\ell=3$) & $7.86855\times 10^{-8}$ & $2.26191 \cdot 10^{-7}$ &  $0.0000171174$  \\
\hline
  2.0 (total) & $0.000100429$ &   $0.000105868$ &  $0.00368425$ \\
  \hline
  2.5 ($\ell=0$) & $0.0000917173$ &  $-$ & $-$  \\
  2.5 ($\ell=1$) & $0.0000349544$ &  $0.00014303$ &  $0.00415494$  \\
  2.5 ($\ell=2$) & $3.25656\times 10^{-6}$ &  $0.0000103005$ &  $0.000523304$  \\
  2.5 ($\ell=3$) & $2.11083\times 10^{-7}$ &  $5.95456 \cdot 10^{-7}$ &  $0.0000387939$  \\
\hline
  2.5 (total) & $0.000130151$ &  $0.000153958$ &  $0.00471937$  \\
\hline
  3.0 ($\ell=0$) & $0.000187681$ &  $-$ & $-$  \\
  3.0 ($\ell=1$) & $0.0000495232$ &  $0.000198045$ &  $0.00503314$  \\
  3.0 ($\ell=2$) & $5.91951\times 10^{-6}$ &  $0.0000182571$ &  $0.000818673$  \\
  3.0 ($\ell=3$) & $5.08632\times 10^{-7}$ &  $1.40793 \cdot 10^{-6}$ &  $0.0000803314$  \\
  3.0 ($\ell=4$) & $3.62776\times 10^{-8}$ &  $9.41478 \cdot 10^{-8}$ &  $6.13981\times 10^{-6}$  \\
\hline
  3.0 (total) & $0.000243671$ &  $0.00021781$ &  $0.00593869$  \\
\hline
  3.5 ($\ell=0$) & $0.000311346$ &  $-$ & $-$  \\
  3.5 ($\ell=1$) & $0.0000676081$ &  $0.000262605$ &  $0.00594792$  \\
  3.5 ($\ell=2$) & $9.96383\times 10^{-6}$ &  $0.0000299928$ &  $0.00121859$  \\
  3.5 ($\ell=3$) & $1.09622\times 10^{-6}$ &  $2.97934 \cdot 10^{-6}$ &  $0.000151374$  \\
  3.5 ($\ell=4$) & $1.00719\times 10^{-7}$ &  $2.57613 \cdot 10^{-7}$ &  $0.0000149082$  \\
\hline
  3.5 (total) & $0.000390123$ &  $0.000295855$ &  $0.00733407$  \\
  \hline
  4.0 ($\ell=0$) & $0.000361305$ &  $-$ & $-$  \\
  4.0 ($\ell=1$) & $0.0000885013$ &  $0.000331921$ &  $0.0068709$  \\
  4.0 ($\ell=2$) & $0.000015623$ &  $0.0000459585$ &  $0.00168665$ \\
  4.0 ($\ell=3$) & $2.13303\times 10^{-6}$ &  $5.69756 \cdot 10^{-6}$ &   $0.000261747$   \\
  4.0 ($\ell=4$) & $2.44477\times 10^{-7}$ &  $6.1674 \cdot 10^{-7}$ &  $0.0000321739$ \\
\hline
  4.0 (total) & $0.000467832$ & $0.000384255$ &  $0.00885492$ \\
 \hline
\end{tabular}
\end{table}

  The semiclassical BH evaporation can be altogether considered as consisting of four 
  stages (certainly without sharp boundaries between them), depending on the BH mass $M$ 
  and, on the last stage, on the deformation parameter $a$:
\begin{enumerate}
\item
 	  $M \gtrsim 10^{17}$\,g, when the evaporation is very slow and mainly involves 
         massless particles. This stage for an initially large  enough BH is much longer than 
         the others, and since $a \ll M$, the BH lifetime is evidently almost the same as for 
         a Schwarzschild BH, given by \eqs (27), (28) of \cite{Page:1976df}.
\item
         $10^{17}\,{\rm g} \gtrsim M \gtrsim 5\cdot 10^{14}\,{\rm g}$.
         Now electrons and positrons are emitted ultrarelativistically and at approximately 
         the same rate for each helicity as neutrinos, but the BH is still large enough for
 	   neglecting the emission of heavier particles \cite{Kanti:2004nr,Kanti:2010mk}.
\item
	   $5\cdot 10^{14}\,{\rm g} \gtrsim M \gtrsim 10^3$\,g (very roughly).        
         It is now necessary to take into account the emission of heavier particles of the 
	   Standard Model. Meanwhile, as before, we have to assume $a \ll M$, hence again 
	   the evaporation rate is actually the same as in the Schwarzschild regime.        
\item
         $M \lesssim 10^3$\,g. Rapid evaporation of tiny BHs approaching the Planck mass 
	   and radius. From some time instant, we can already assume $a \sim M$, so that the 
	   grey-body factors substantially enlarge the intensity of Hawking radiation and accelerate 
	   the BH evaporation. Moreover, the BH temperature is large enough to strongly exceed 
	   the masses of all Standard Model particles, hence all of them are emitted 
	   ultrarelativistically and at approximately the same rate as neutrinos, electrons and 
 	   positrons, thus substantially increasing the corresponding multiplicity factors.
	   This stage is certainly much shorter than any of the previous ones and is even more 
	   shortened by the larger grey-body factors at $a >0$.
\end{enumerate}
  Ultimately, closer to the Planck parameters ($M_{\rm Pl}\approx 10^{-5}$ g), we lose a 
  reason to apply the semiclassical approximation and approach the realm of full quantum gravity.

  Table I shows that when the deformation parameter $a$ is turned on, the intensity of Hawking
  radiation is greatly enhanced. For $a/M < 1$, the total emission of scalar, Dirac and Maxwell
  particles grows by tens of percent, and at larger $a/M$ the emission can be a few times larger
  than from a Schwarzschild BH. This effect can be explained by the properties of the effective
  potentials shown in Figs.\,\ref{figVscal}--\ref{figVDir}: the potential barrier becomes much lower
  as $a$ increases, hence fewer particles are reflected by this barrier back to the event horizon.

  It is worth noting that the WKB results for the energy emission rate given in Table I for the lowest
  $\ell$ differ from the accurate values found by numerical integration in the seminal paper by 
  Page \cite{Page:1976df} (see Table I there). 
  As mentioned above, such a comparison can 
  be done if we restrict our data to the sector of particles considered by Page, i.e., take the 
  multiplicity factor in \eqref{multiDir} as $N_\ell=8\ell$. In this case, for $\ell = 1$, the Dirac 
  field, Page's result for $dE/dt$ is $1.575 \cdot 10^{-4}$, while we would have 
  $1.596 \cdot 10^{-4}$. For the Maxwell field, Page's rate of $0.330 \cdot 10^{-4}$ versus 
   our WKB result of $0.335 \cdot 10^{-4}$.  Thus the relative error is about $1\%$ in the   
   Schwarzschild limit, which is quite negligible, considering that the effect 
   (deviation from the Schwarzschild limit) reaches tens or even  hundreds of percent.

  Figures \ref{ScalRate0}--\ref{fig52} show that the maximum of the energy emission rate 
  per frequency unit is slightly shifted towards smaller $\omega M$. As $a$ increases, the 
  contribution of higher multipoles $\ell$ into the total flux becomes larger, so that summing 
  over the first three multipoles becomes insufficient. Therefore, we considered contributions 
  from the first five multipoles for all
  values of $a$, but not all data for higher multipoles are given in Table I. As shown by the 
  examples in Figs.\,\ref{ScalRate0}--\ref{fig52}, higher multipoles can be neglected. Even for 
  relatively small values of $a$ (e.g., $a = 0.5$), for which the fundamental quasinormal modes
  deviate from their Schwarzschild values by less than $2\%$, the deviation in the intensity of
  Hawking radiation is an order of magnitude larger.
  
  As to numerical estimates, using \eq \rf{M-dot} and the data from Table I, it is easy to obtain, 
  for example, that a Schwarzschild BH with the initial mass of $10^{15}$\,g will have a lifetime 
  of  $\sim 9\times 10^{17}$ seconds in agreement with  
  the well-known data on primordial BH survival up to the present epoch. On the other hand,
  a Schwarzschild BH of 1 g $\sim 10^5\,M_{\rm Pl}$ should evaporate within $\sim 10^{-27}$\,s. 
  One can conclude that the effect under study can further shorten the evaporation time but only
  at the very end of this process, enhancing the energy release in several times. A precise calculation
  of this process is complicated by the fact that for given $a$ the ratio $a/M$ rapidly grows 
  as the BH mass $M$ decreases, so that the coefficient $\alpha_0$ in \eq \rf{M-dot} grows more 
  and more as the BH mass approaches the Planckian value.

\section{Quasinormal modes: Overtones outburst}

  In addition to quantum Hawking radiation, there is a classical characteristic that significantly 
  depends  on the near-horizon geometry: it is the first few overtones of the 
  {\it quasinormal modes} 
  (QNM) of BHs \cite{Kokkotas:1999bd,Konoplya:2011qq}. QNM represent proper 
  oscillations under specific boundary conditions: purely incoming waves at the event horizon 
  and purely outgoing ones at infinity \cite{Konoplya:2011qq}. These boundary conditions are 
  different from the scattering ones used here while studying the grey-body factor since when 
  discussing  classical radiation processes, we imply that the event horizon corresponds to 
  complete absorption. 

  To obtain the precise values of QNM with $n>\ell$, where $n$ is the overtone 
  number, we use the Frobenius method, applied by Leaver for QNM calculation in BHs 
  \cite{Leaver:1985ax}. The wavelike equation always exhibits a regular singularity at the horizon 
  $r=r_0$ and an irregular singularity at spatial infinity $r=\infty$. To address that, we introduce 
  a new function:
\begin{equation}\label{reg}
		\Psi(r)= P (r, \omega) \left(1-\frac{r_0}{r}\right)^{-\imo\omega/F'(r_0)}y(r),
\end{equation}
  Here, the factor $P$ is chosen so that $y(r)$ is regular for $r_0\leq r<\infty$, ensuring that 
  $\Psi(r)$ corresponds to a purely outgoing wave at spatial infinity and a purely ingoing wave 
  at the horizon. Then we present $y(r)$ in the form of a Frobenius series:
\begin{equation}\label{Frobenius}
		y(r)=\sum_{k=0}^{\infty}a_k\left(1-\frac{r_0}{r}\right)^k.
\end{equation}

  While the fundamental mode is determined by the geometry near the peak of the effective 
  potential and is thereby insensitive to near-horizon deformations, overtones are highly sensitive 
  even to relatively small such deformations 
  \cite{Konoplya:2023hqb, Konoplya:2022pbc, Konoplya:2022hll} and thus may provide 
  information on quantum corrections to the geometry. The first several overtones are essential 
  for reproducing the early stage of the ringdown phase \cite{Isi:2019aib}, while the fundamental
  mode alone allows one to reproduce only the very end of this process. Although the fundamental 
  mode has been extensively studied in \cite{Konoplya:2019xmn, Saleh:2014uca, Saleh:2016pke}, 
  no such analysis for overtones has been done so far.

\begin{figure}
\resizebox{\linewidth}{!}{\includegraphics*{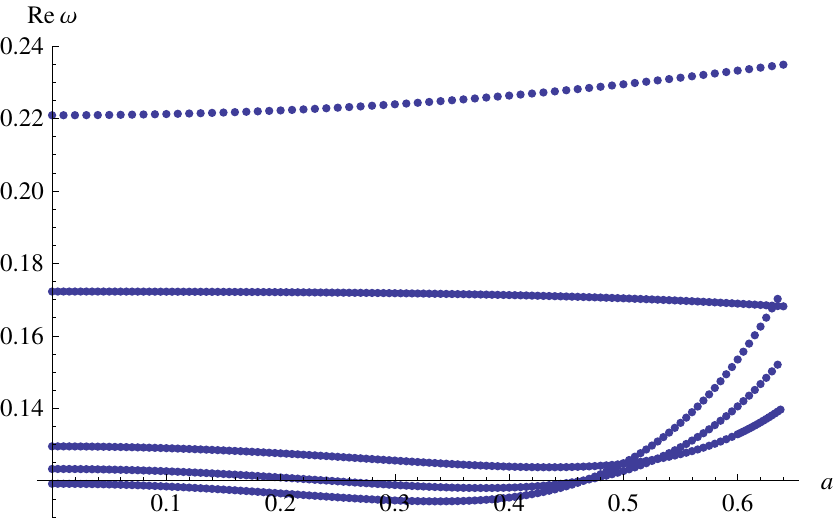}}
\caption{$\re{\omega}$ as a function of the quantum correction parameter $a$ for scalar field
	 perturbations; $r_{H}=1$, $\ell=0$, for various overtones: $n=0, 1, 5, 7, 9$ from top to bottom. 
	 }\label{figQNM1}
\end{figure}

\begin{figure}
\resizebox{\linewidth}{!}{\includegraphics*{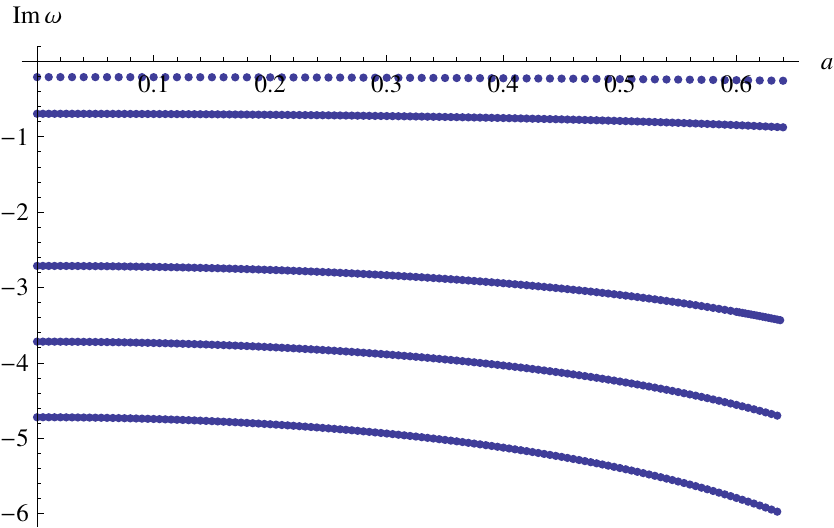}}
\caption{$\im{\omega}$ as a function of the quantum correction parameter $a$ for scalar field 
	perturbations; $r_{H}=1$, $\ell=0$, for various overtones: $n=0, 1, 5, 7, 9$ from bottom to top.
	}\label{figQNM2}
\end{figure}

  Here we observe that there is indeed an outburst of overtones when the parameter $a$ is     
  sufficiently large, i.e., the overtones deviate from the Schwarzschild values at an increasingly 
  higher rate. Conversely, for sufficiently small $a$, the fundamental mode and the overtones 
  deviate from their Schwarzschild values at approximately the same rate.  Indeed, as can be seen 
  in Fig.\,\ref{figQNM1}, the oscillation frequency $\re{\omega}$ (rather than the damping rate 
  $\im{\omega}$ in Fig.\,\ref{figQNM2} that asymptotically scales with the overtone number $n$) 
  differs from the Schwarschild one by a few percents for $n=0$ but by a few tens of percent 
  for $n=5$ and higher. This behavior can be easily explained according to \cite{Konoplya:2022pbc}: 
  as the Hawking temperature is the same as in the Schwarzschild case, the outburst of overtones 
  develops relatively slowly, unlike those BHs whose temperature is different from the 
  Schwarzschild one.

It is worth noting that the two different spectral problems considered here are, in fact, connected: the gray-body factors can be accurately reproduced from quasinormal modes through the correspondence found in \cite{Konoplya:2024lir,Konoplya:2024vuj} for finite \(\ell\), and exactly in the eikonal limit as \(\ell \rightarrow \infty\). The precise values of the quasinormal modes shown in Figs. 13 and 14 are available from the authors upon request.


\section{Conclusions}

  Our analysis has revealed that although the classical radiation, characterized by quasinormal
  modes, changes only slightly due to the quantum deformation parameter \cite{Konoplya:2019xmn},
  the Hawking radiation is significantly enhanced by the quantum corrections. Remarkably, this effect
  occurs even though the Hawking temperature of a deformed black hole is almost the same as
  that of a Schwarzschild one. The enhancement of Hawking radiation intensity is substantial, 
  exceeding tens or even hundreds of percent deviation from the Schwarzschild limit,
  due to considerably increased grey-body factors that grow because of much lower potential
  barriers in quantum-corrected BHs, which allow a much larger fraction of the initial radiation
  to reach the infinity. 

  We have focused on the emission of massless fields since any masses of known particles are
  many orders of magnitude smaller than the energy scale of the phenomenon under study.
  However, the problem of massive field emission at BH evaporation is significant
  \cite{Page:1976ki,Page:1976ki}, and we can notice that although the field mass should in general
  diminish the emission rate, which agrees with the observation made in \cite{Kanti:2010mk},
  the corresponding  effective potentials become lower when the deformation parameter $a$ 
  is taken into consideration, hence, similarly to massless particles, massive ones should 
  be emitted more intensively around a quantum-deformed BH as compared to a Schwarzschild one.
  To estimate the emission rates of massive particles, the WKB formula used here would not 
  provide a good approximation because the effective potential has three turning points in this case.

  In addition to the accelerated BH evaporation at its latest stage, one more consequence 
  of the BH quantum correction, observable in principle but also hardly expected to be detectable 
  in the foreseeable future is the predicted outburst of QNM overtones, which are sensitive to the least changes of the near-horizon geometry.

For $M \lesssim 10^{15}$\, g, primordial black hole evaporation is mostly constrained by big bang nucleosynthesis \cite{Carr:2020gox}.
Therefore, the quantum corrections considered here could be one of the mechanisms (in addition to those considered in \cite{Franciolini:2023osw}) that make black holes evaporate faster and, thereby, evade big bang nucleosynthesis constraints on primordial black holes in dark matter. However, the quantum correction considered here will influence only the latest stages of evaporation, keeping the classical estimations for the total lifetime of primordial black holes unchanged.

An intriguing question that had previously escaped our consideration concerns the impact of 
back reaction of quantized fields on the black hole background. Initially, the Kazakov-Solodukhin metric is derived by accounting for spherically symmetric quantum fluctuations of the metric 
and the matter fields nonperturbatively \cite{Kazakov:1993ha}. Consequently, only nonspherical
fluctuations of the fields are disregarded in the aforementioned approach. However, it is widely acknowledged that during the final stages of evaporation, the black hole's angular momentum 
is emitted \cite{Kanti:2004nr}, rendering the black hole highly spherically symmetric. 
This justifies the omission of non-spherical back-reaction effects.

\acknowledgments
  {This work was supported by the RUDN University Project No. FSSF-2023-0003.
 

\end{document}